**ORIGINAL ARTICLE**

# *JWST* Imaging of the Closest Globular Clusters – I. Possible Infrared Excess Among White Dwarfs in NGC 6397[†]

L. R. Bedin*[1] | D. Nardiello[1,2] | M. Salaris[3] | M. Libralato[1] | P. Bergeron[4] | A. J. Burgasser[5] | D. Apai[6,7] | M. Griggio[1,8,9] | M. Scalco[1,8] | J. Anderson[9] | R. Gerasimov[10] | A. Bellini[9]

[1]Istituto Nazionale di Astrofisica, Osservatorio Astronomico di Padova, Vicolo dell'Osservatorio 5, Padova, IT-35122, Italy

[2]Università di Padova, Dipartimento di Fisisca e Astronomia, Vicolo dell'Osservatorio 3, Padova, IT-35122, Italy

[3]Astrophysics Research Institute, Liverpool John Moores University, 146 Brownlow Hill, Liverpool L3 5RF, UK

[4]Département de Physique, Université de Montréal, C.P. 6128, Succ. Centre-Ville, Montréal, QC H3C 3J7, Canada

[5]Department of Astronomy & Astrophysics, University of California San Diego, La Jolla, CA 92093, USA

[6]Department of Astronomy and Steward Observatory, The University of Arizona, 933 N. Cherry Avenue, Tucson, AZ 85721, USA

[7]Lunar and Planetary Laboratory, The University of Arizona, 1629 E. University Blvd., Tucson, AZ 85721, USA

[8]Università di Ferrara, Dipartimento di Fisica, Via Giuseppe Saragat 1, I-44122, Ferrara, Italy

[9]Space Telescope Science Institute, 3800 San Martin Drive, Baltimore, MD 21218, USA

[10]Department of Physics and Astronomy, University of Notre Dame, Notre Dame, Nieuwland, Science Hall, Indiana, 46556, USA

**Correspondence**
*E-mails: luigi.bedin@inaf.it

We present *James Webb Space Telescope* observations of the globular cluster NGC 6397 and use them to extend to infrared wavelengths the characterization of the cluster's entire white dwarf (WD) cooling sequence (CS). The data allows us to probe fundamental astrophysical WD properties and to search for evidence in their colors for (or against) the existence of ancient planetary systems. The existing archival *Hubble Space Telescope* imaging data obtained ∼18 years ago reach ultra-deep optical magnitudes ($V$ ∼31) and allow us to derive a near-perfect separation between field and cluster members. We detect an apparent split in the lower part of the WD CS of NGC 6397. The red part of the WD CS, containing about 25% of the total, exhibits significant IR-excess of up to $\Delta m_{\rm F322W2} \sim 0.5$ mag. These infrared excesses require both theoretical and observational follow-ups to confirm their veracity and to ascertain their true nature.

[†]Based on observations with the NASA/ESA *James Webb Space Telescope*, obtained at the Space Telescope Science Institute, which is operated by AURA, Inc., under NASA contract NAS 5-26555, under GO-1979.

[0]**Abbreviations:** *JWST*, James Webb Space Telescope; *HST*, Hubble Space Telescope; ⊙, Sun/Solar

## 1 | INTRODUCTION

Globular clusters (GCs) are the oldest objects in the Universe for which accurate ages can be determined. They are ideal laboratories because, to a first approximation, they consist of stars of the same age, distance, and chemical composition. Stellar color-magnitude diagrams (CMDs) of GCs are important tools for stellar astrophysics and dynamics. Accurate characterization of GC populations requires understanding and eliminating field contamination in the observed regions. Our program allows us to test *JWST*'s astrometric capabilities for this purpose (Griggio, Nardiello, & Bedin, 2023). By leveraging existing high-resolution *HST* images collected up to ~20 yrs ago, we can use proper-motion measurements to establish cluster membership for the faintest stars and WDs in these systems. This article describes the first scientific exploration of imaging data collected with *James Webb Space Telescope (JWST)* program GO-1979 (PI: Bedin), which aims to measure high-precision infrared photometry and astrometry of the faintest objects in the two closest Galactic GCs, Messier 4 ($d$=1.83±0.03 kpc) and NGC 6397 ($d$=2.46±0.06 kpc Baumgardt & Vasiliev, 2021). The main goals of this program include exploring multiple stellar populations among the low-mass stars in these GCs (Marino et al., 2008; Milone et al., 2012a), including substellar populations; examining the clusters' internal kinematics which provides important information about the formation of the clusters' multiple populations (see, e.g. Cordoni et al., 2020; Martens et al., 2023; Tiongco, Vesperini, & Varri, 2019), including dispersion and rotation as a function of stellar mass (Scalco et al., 2023), and examining the outer regions of the cluster, and the Galactic field populations.

Here, we present analyses of the entire white dwarf (WD) cooling sequence (CS) of NGC 6397. This cluster hosts multiple populations disclosed both photometrically (Milone et al., 2012b) and spectroscopically (see, e.g., Gratton et al., 2001), an iron abundance [Fe/H]~ −2.0 (Gratton et al., 2001), and main sequence turn off ages in the range ~12.5-13 Gyr (Correnti, Gennaro, Kalirai, Cohen, & Brown, 2018; VandenBerg, Brogaard, Leaman, & Casagrande, 2013).

Our investigation of WDs extends to the infrared (IR) band the study already done in the optical with photometry collected with the *Hubble Space Telescope* (*HST,* Anderson, King, et al., 2008; Hansen et al., 2007; Richer et al., 2006; Torres, García-Berro, Althaus, & Camisassa, 2015) that was tuned towards the determination of the cluster age from the cooling sequence, and has provided ages in the range ~ 11.5-13.0 Gyr.

Many WDs have shown infrared excess in their spectra (e.g., Reach et al., 2005; Su et al., 2007), and recent estimate is that 15% of young WDs possess a debris disk (Chu et al., 2011). Infrared observations of WDs in GCs can reveal the presence of debris disks in these systems. Indeed, it is commonly assumed that metal-enrichment in DZ WDs (Metal lines present in atmosphere) is explained as accretion of tidally-disrupted planetesimals (Swan et al., 2023, e.g.).

The extended wavelength range in the here-presented *JWST* data allows us to explore the evolution and the fundamental astrophysics of WDs in this GC; in particular, examining evidence for (or against) the existence of ancient planetary systems around hotter WDs through the presence (or absence) of IR photometric excess.

## 2 | OBSERVATIONS

### 2.1 | *JWST* images

We examined images collected with the *JWST* Near Infrared Camera (NIRCam, Rieke et al., 2023) of NGC 6397 taken under program GO-1979. Observations were conducted simultaneously with the Short Wavelength (SW) and Long Wavelength (LW) channels on 2023 March 14 (epoch ~2023.2). We used the 6-point `FULLBOX` primary pattern with a `2-POINT-LARGE-WITH-NIRISS` sub-pixel dither pattern. At each of the resulting twelve pointings, a single image in both channels was collected with the readout pattern `MEDIUM8` (6 groups), for an effective exposure time per image of 622.733 s. We used ultra-wide filters for both NIRCam channels: F150W2 in the SW channel and F322W2 in the LW channel. Left-panel of Figure 1 shows a Digital Sky Survey 2 InfraRed[1] image of the field around our NIRCam pointing, which is centered at $(\alpha;\delta)$ = (265°.275; −53°.756), at an average angular distance from the center of the cluster of about 6 arcmin, ranging between 2.7 arcmin and 9.2 arcmin. The central panel of Figure 1 shows the F150W2 stacked image for the entire NIRCam field of view (FoV), while the right panel shows a zoom-in of $1' \times 1'$ for a relatively "sparse" region highlighting both the high density of stars and abundant background of resolved extragalactic sources.

We independently ran stages 1 and 2 of the *JWST* calibration pipeline (Bushouse et al., 2023) to obtain the level-2b `_cal` images. In the stage 1 pipeline, we used all default parameters except for the ramp fit, for which we used the `frame zero` (the first frame of each integration) to measure pixels that are saturated in the first group up to the ramp. This option increases the dynamical range of our data by over two magnitudes into the nominally saturated intensity regime. We ran the stage 2 pipeline with all default parameters. We then modified the `_cal` images by converting the values of the pixels from MJy sr$^{-1}$ into counts by using the FITS file header keywords `PHOTMJSR` (conversion factor from MJy sr$^{-1}$ to DN s$^{-1}$) and `XPOSURE` (the effective exposure time of each pixel).

---

[1] Plate ID: A30L collected 1988/08/20; for details see http://archive.eso.org/cms/tools-documentation/-the-eso-st-ecf-digitized-sky-survey-application.html

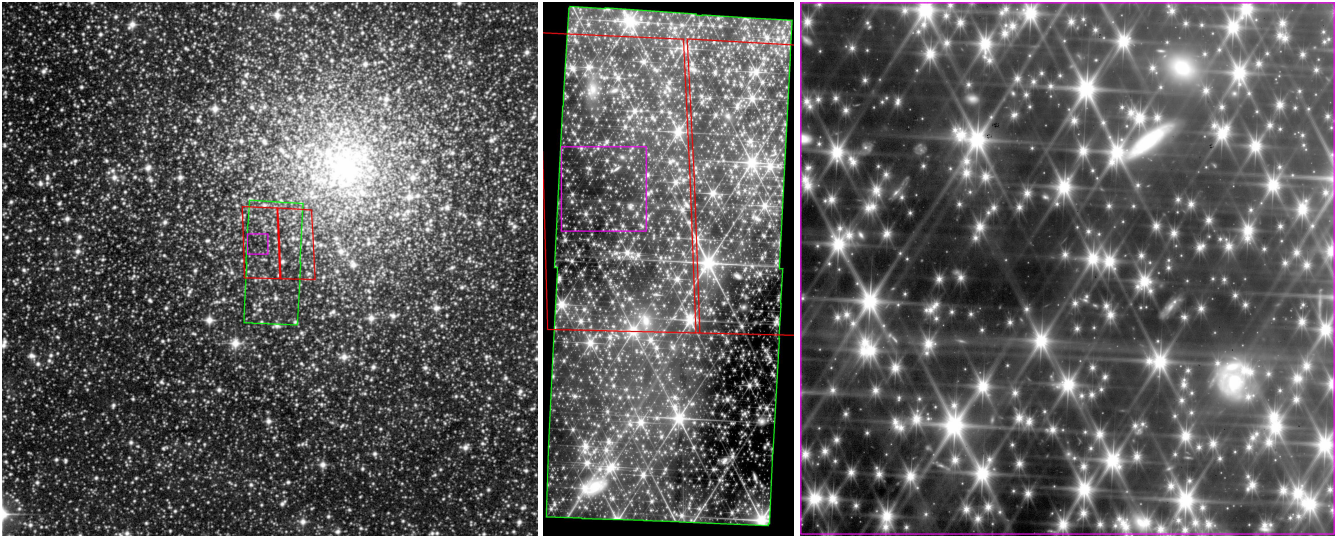

**FIGURE 1** *(Left:)* A $25' \times 25'$ infrared image from the Digital Sky Survey 2 centered on our NIRCam field of NGC 6397 for *JWST* program GO-1979 (green box). The image is aligned with North up and East toward the left. The region indicated in red is the archival *HST* deep field from programs GO-10424 and GO-11633. *(Middle:)* The entire NIRCam primary field in filter F150W2. *(Right:)* A zoom-in on a dark sub-region (of $\sim 1' \times 1'$) in the NIRCam image (magenta boxes in the left and center frames), at a scale of $60''$ where individual pixels are visible.

We also flagged unusable pixels by using the data quality flags contained in the `_cal` data cube.

## 2.2 | *HST* images

We planned our NIRCam observations to overlap as much as possible with archival data collected by *HST* with the Advanced Camera for Surveys (ACS) Wide Field Channel (WFC) in broad-band filters F606W and F814W, obtained through programs GO-10424 (PI: Richer) and GO-11633 (PI: Rich). Given the relative proximity and low reddening of NGC 6397, these observations are the deepest optical *HST* observations available of any GC. The first *HST* epoch was taken in $\sim$ 2005.2 and consisted of 252 F814W images with exposure times between 616 s and 804 s, and 126 F606W images with exposure times between 630 s and 769 s. The second *HST* epoch was taken in $\sim$ 2010.2 and consisted of 18 F814W images obtained with exposure times between 1280 s and 1405 s. In the latter program, short-exposure images ($\lesssim$40 s) were also taken, which were useful in measuring a complete sample of bright stars that were masked during the data reduction of the deep exposures (see Anderson, Sarajedini, et al. 2008 for details on these data sets).

## 3 | DATA REDUCTION

All images obtained with NIRCam were reduced with software tools and methods described in detail in Papers I, II, and III of the series *"Photometry and astrometry with JWST"* (Griggio et al., 2023; Nardiello et al., 2022; Nardiello, Bedin, et al., 2023), and applied with success in the recent Nardiello, Griggio, and Bedin (2023) study of brown dwarfs in 47 Tucanæ. Our methodology consists of a *first-* and *second-pass* photometry as described in Anderson, King, et al. (2008). In this section, we briefly describe the approach, but refer readers to the Anderson reference for a comprehensive description.

The *first-pass* photometry essentially collects all the information on PSFs and coordinate transformations from each image into the common reference frame. In our case, we adopted *Gaia* DR3 bright members as a reference, after having transformed their positions to the epoch of the data collection (Nardiello et al., 2022; Nardiello, Bedin, et al., 2023) and applying our current best geometric distortion correction (Griggio et al., 2023).

In the *second-pass* photometry, we used a modified version of the code KS2 developed by Jay Anderson and described in Bellini et al. (2017), Scalco et al. (2021), and references therein. In this step, we extracted positions and fluxes using both the PSFs and the transformations (from coordinates of individual-images to the master frame) as derived in the first-pass photometry (see Griggio et al., 2023; Nardiello et al.,

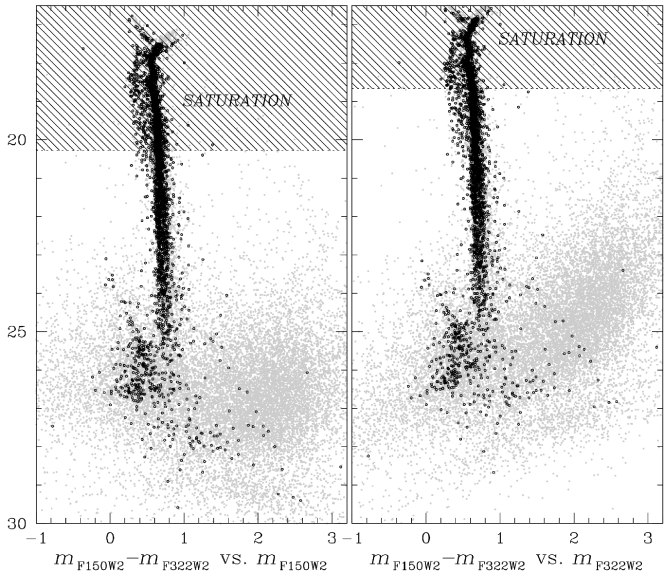

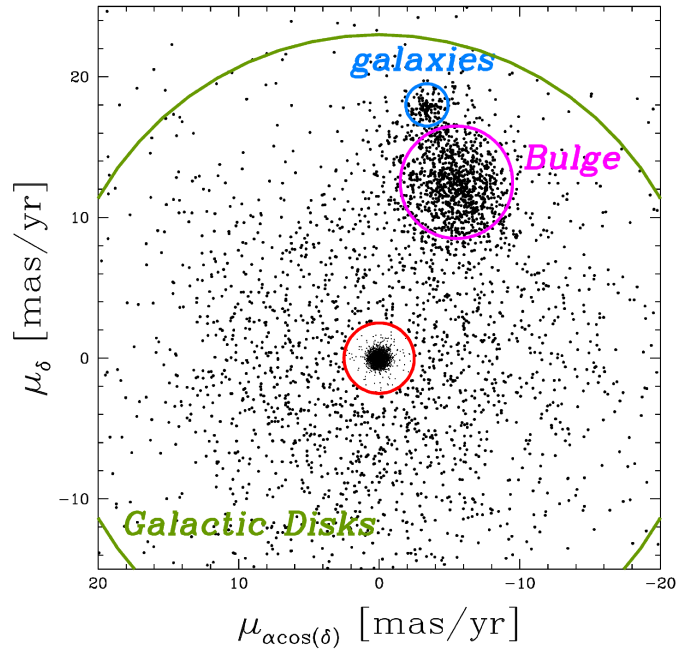

**FIGURE 2** The two CMDs of NGC 6397 possible with the *JWST* NIRCam F150W2 and F322W2 filters. All local maxima detected within the entire NIRCam field of view are shown as grey dots, most being artifacts or background galaxies. Sources passing all quality criteria are shown as black circles (see text). The dashed regions indicate the on-set of saturation in the normal NIRCam up-the-ramp sampling, but measured here through the use of the `frame zero` option.

**FIGURE 3** Vector point diagram for sources that passed all selection criteria based on `KS2` diagnostics. The inner red circle corresponds to motions within 2.5 mas/yr of the bulk of point sources in the NIRCam field (defined as cluster members; see text). We also indicate locations for the bulk of extragalactic sources and the external portion of the Galactic Bulge (magenta circle). The wider green circle encompasses nearby Galactic thin disk and thick disk stars, as well as poorly measured members.

2022). `KS2` makes use of all pixels from all images simultaneously, making it particularly suitable to obtain deep photometry for sources too faint to be detected in individual images. Along with fluxes and positions, `KS2` produces a number of quality diagnostics, including the root mean square (RMS) error in brightness (in magnitudes), the RMS error in positions, the PSF quality of fit (q), and a "stellarity index" that describes how well a given source's shape resembles the PSF (aka `RADXS`, see Bedin et al. 2008).

In the wide-passband filters deployed, the spectral energy distribution (SED) of a source do modifies the shape of the PSF and hence it could potentially impacts the observed positions. Indeed this was the case for *HST* optical detectors (e.g., Bellini, Anderson, & Bedin, 2011). However, while discernible for high-signal-to-noise-ratio point sources, these offsets are completely negligible at the faint magnitudes of interest in the present work, for which random errors of $\sim 0.5$ pixel dominate (NIRCam pixel scale is $\sim 31.2$ mas).

We conducted an analogous reduction procedure for the *HST* data sets, performing the same *first-* and *second-pass photometry*, employing for the second-pass the original version of `KS2` that was developed specifically for *HST* cameras Anderson, King, et al. (2008). We followed procedures described in detail in Bedin et al. (2023) and in references therein.

Having measured the photometry and astrometry in data units (digital numbers and pixels), we calibrated the *JWST* and *HST* photometry to the VEGA-magnitude system following standard procedures (e.g., Bedin et al., 2005a), while the astrometric positions were calibrated to the International Celestial Reference System (ICRS) frame using *Gaia* DR3 data for sources in the observed fields, extrapolated to the reference epoch of the *JWST* dataset (i.e., 2023.2).

Figure 2 shows the two CMDs based on our F150W2 and F322W2 photometry for the sources in the entire NIRCam field of view (FoV). All local maxima are shown in grey while sources selected with the applied `KS2` quality flags (such as `rmsSKY`, `RADXS`, `q`, `etc`, see Anderson, Sarajedini, et al. 2008) are marked in black (see appendix A for a description on the selections). The CMDs extend over 12 magnitudes in brightness thanks to the NIRCam `frame zero` sampling option.

## 3.1 | Proper Motions

Proper motions (PMs) between the *JWST* observations and earliest archival *HST* data were computed as displacements between the two epochs divided by the temporal baseline, approximately 18 yr. Displacements were computed relative to the group of detected point sources that moved the least, identified as the cluster members. NGC 6397 stars have an internal dispersion of about 5 km s$^{-1}$ (Vasiliev & Baumgardt, 2021), which at its distance of ~2.5 kpc corresponds to a PM dispersion of less than 0.5 mas yr$^{-1}$, or less than 10 mas in 18 years, about one third of a pixel. This value should be regarded as an upper limit for the dispersion of this sample, as the outskirts of the cluster where our NIRCam field is placed have a lower dispersion than the cluster overall. These internal dispersion limits are considerably smaller than the random proper motion uncertainties at the faint magnitudes of the bulk of the WDs (much less than 0.5 mas yr$^{-1}$ for the expected internal motions, vs. ~2 mas yr$^{-1}$ for the observed random errors). Figure 3 shows the vector-point diagram (VPD) for the point sources with proper motion measurements. The core of this plot has 1D-dispersions of ~0.5 mas yr$^{-1}$. To balance exclusion of contaminating field stars with inclusion of faint cluster stars with larger PM errors, we set a cluster membership threshold of $\mu \leq 2.5$ mas yr$^{-1}$ (or about 5 times the expected maximum internal PMs dispersion).

For comparison, our VPD in Figure 3 highlights the expected locations for extragalactic objects and the external part of the Galactic Bulge at a typical distance of $\mathcal{R} \sim 7.4$ kpc.[2] NGC 6397 is located about 25° from the Galactic Center, at $(\ell; b) \simeq (-22°; -12°)$, and the *HST* and *JWST* sight-line probes the external part of the Galactic Bulge, resulting in a sizable component of the observed field objects. Simulations such as those by Libralato et al. (2018) show that the VPD maps a continuity of relative proper motion along the Galactic plane from extragalactic sources, to the external parts of the Galactic Bulge to the (more broadly dispersed) Galactic thin-disk and thick-disk, including the relatively nearby foreground stars.

Finally, we note that in addition to separating field foreground and background stars from cluster members, PMs are a great diagnostic for removing artifacts and poorly measured sources. Indeed, PMs are more effective than relying on the RADXS diagnostic, which KS2 failed to compute for a large fraction of the local maxima in our NIRCam data (about 20%).

Fig. 4 shows the CMD –focused on the WD loci– for all unsaturated sources with measurable PMs in the overlapping *HST* and *JWST* fields. We also show the total PM $\mu = \sqrt{\mu^2_{\alpha\cos\delta} + \mu^2_\delta}$ as a function of magnitude $m_{F150W2}$, highlighting our membership criterion of $\mu < 2.5$ mas yr$^{-1}$. The final proper-motion cleaned CMD shows that the bulk of background sources (at $m_{F150W2} - m_{F322W2} \approx 0.7$) are removed, leaving the bottom of the stellar MS and the WD CS.

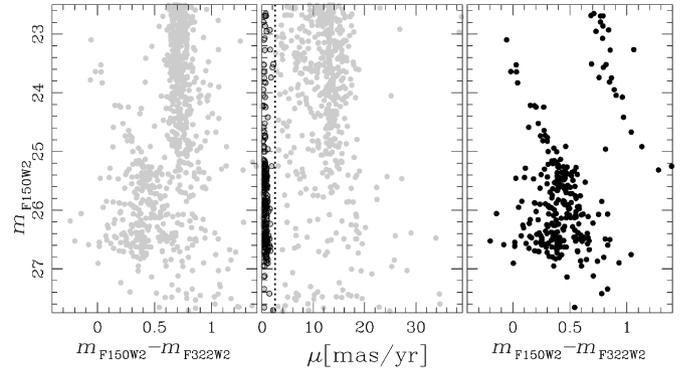

**FIGURE 4** Proper-motion cleaned sample of NGC 6397 cluster members for those sources that fall in both *JWST* NIRCam and *HST* ACS/WFC fields of view (Fig 1 ). The *left* panel shows the CMD of unsaturated point sources in the WD CS region with measured proper motions, and which passed all KS2 selection criteria (grey dots, see text). The *middle* panel shows total relative proper motions to the cluster median versus F150W2 magnitude, with the ≤2.5 mas/yr threshold for cluster members indicated as a vertical dotted line. Members are encircled in black. The *right* panel shows the CMD for proper motion-selected members, focusing on the WD CS region.

## 3.2 | Artificial Star Tests and Sample Completeness

Artificial star tests (ASTs) are an important step in assessing the reliability of point source photometry and sample completeness. We generated 100,000 artificial stars within the NIRCam FoV (~70,000 of which were within the field in common with ACS, see Fig. 1 ), placing them at consistent positions within each individual image using the corresponding position-dependent PSFs for each camera and filter combination. We assumed a uniform distribution in F322W2 magnitude. For the other three filters (x=F606W, F814W, and F150W2), we determined corresponding magnitudes based on the WD CS fiducial lines for the corresponding CMDs (x−F322W2 vs. F150W2). We then used the appropriate version of the KS2 software used to find and measure real stars to find and measure artificial stars in the *HST*/ACS/WFC and *JWST*/NIRCAM images. We considered artificial stars "recovered " when their measured positions and magnitudes

---

[2] $\mathcal{R} = \mathcal{R}_\circ \cos\ell \cos b$ (Bedin, Piotto, King, & Anderson, 2003); assuming $\mathcal{R}_\circ = 8.2$ kpc (GRAVITY Collaboration et al., 2019).

where within 0.5 pixel and 1 magnitude of their assigned values, respectively.

ASTs have several important uses for assessing the fidelity and completeness of our sample.

First, they provided a *fundamental* correction, which is used to remove the so-called "migration" effect, a systematic offset in magnitude when measuring the faintest sources. These corrections are also called input/output corrections, and are needed to bring the measured magnitudes in line with the true values (see Bedin et al. 2009 for a more extensive description of these effects). We found these corrections to be negligible (<0.1 mag) for the *JWST* data down to the faintest magnitudes studied, but reached deviations as large as 0.5 mag in the *HST* images (cf. Figs. 2 and 3 in Bedin et al. 2009).

Second, artificial stars were used to estimate random errors. Reliable error estimates are necessary to generate accurate simulated observations, which are in turn a pre-requisite for accurate comparison between models and observations (see Sect. 4.4 simulated CMD). Estimates for photometric errors are computed, in half-magnitude bins, from the dispersion of the measured values minus the inserted values.

Third, artificial stars are used to define the selections based on diagnostic quality parameters generated by KS2, such as rmsSKY, RADXS, q and even the location of the WDs on the CMD (details are given in the Appendex, but more examples on selections are given in Bedin et al. 2008, 2023, 2009). Basically we require that real stars to have the same qualityparameters distribution, as the bulk of the well recovered artificial stars.

Fourth, artificial stars were used to assess "consistency in position" between the *HST* and *JWST* epochs. While by construction artificial stars have no position offset, at faint magnitudes flux noise on top of the PSF shape can result in spurious offsets. This test provides a lower limit on the astrometric measurement errors (∼0.3 mas yr$^{-1}$).

Finally, artificial stars are used to evaluate the fraction of missed stars, and hence the completeness of WDs recovered. Completeness curves also establish how faint we can go before the reliability in sample selection is compromised. Following Bedin et al. (2008, 2023) and reference therein, we adopted as our faint limit the magnitude at which completeness in "good regions" ($c_g$) is 50%. Here, good regions are those in which sources can be found above the local background, and excludes the regions around the bright halos of saturated stars (cf. Bedin et al., 2023). The completeness limit for the *JWST* data is $m_{F150W2} = 27.17$ and $m_{F322W2} = 26.88$, both well below the drop in the number of observed WDs.

# 4 | THE WHITE DWARF COOLING SEQUENCE OF NGC 6397

## 4.1 | Color-Magnitude Diagram and Isochrones

In Figure 5, we show the two most informative CMDs for the WD CS in NGC 6397, namely F150W2−F322W2 versus F150W2 and F814W−F150W2 versus F322W2. These CMDs show only unsaturated stars that pass all the quality selection criteria including proper motion membership. Although these CMDs are less complete, they contain the most reliably characterized stars.
The F814W−F150W2 versus F322W2 CMD disentangles two effects that cause the WD CS to turn blueward at the faint end, as discussed in detail in Sect. 4.3. The F150W2−F322W2 versus F150W2 CMD, on the other hand, highlights an unexpected feature that seems to suggest the presence of IR-excess in a subset of WDs, as discussed in detail in Sect. 4.4.

We compare to these CMDs a representative 13 Gyr DA WD isochrone calculated with the BaSTI-IAC hydrogenatmosphere WD cooling tracks (Salaris et al., 2022), calculated with the Cassisi, Potekhin, Pietrinferni, Catelan, and Salaris (2007) electron conduction opacities (see Salaris et al., 2022, for a discussion on this issue), the initial-final mass relation by Cummings, Kalirai, Tremblay, Ramirez-Ruiz, and Choi (2018) and the progenitor lifetimes from Pietrinferni et al. (2021) for an [α/Fe]=0.4 metal mixture and [Fe/H]=−1.9, appropriate for the cluster (see, e.g. Horta et al., 2020). We assumed a distance modulus $(m-M)_0$=11.96 (2.47 kpc) based on *Gaia* EDR3 parallaxes (Baumgardt & Vasiliev, 2021), $E(B-V)$=0.18 (Gratton et al., 2003), $A_V = 3.1 E(B-V)$, and extinction ratios $A_\lambda/A_V$ from Wang and Chen (2019) for the *JWST*/NIRCAM filters and from Bedin et al. (2005b) for the *HST*/ACS/WFC filters.

## 4.2 | The WD Luminosity Function

The top panel of Fig. 6 shows the differential luminosity function (LF, thin line) for the observed WDs. The LF was determined by counting WD stars within bins of 0.1 magnitude over the interval $m_{F150W2}$ between 23 and 28. We only used unsaturated sources that passed all selection criteria (see Appendix A1), with the exception of the RADXS parameter. We did not make use of RADXS because it was not provided by KS2 for a large fraction (∼20%) of sources, therefore impacting the completeness of the sample. Although this sample is slightly noisier photometrically, it remains robust in terms of membership, and able to follow the LFs well below its peak, which is a magnitude brighter than the $c_g$ =50% limit,

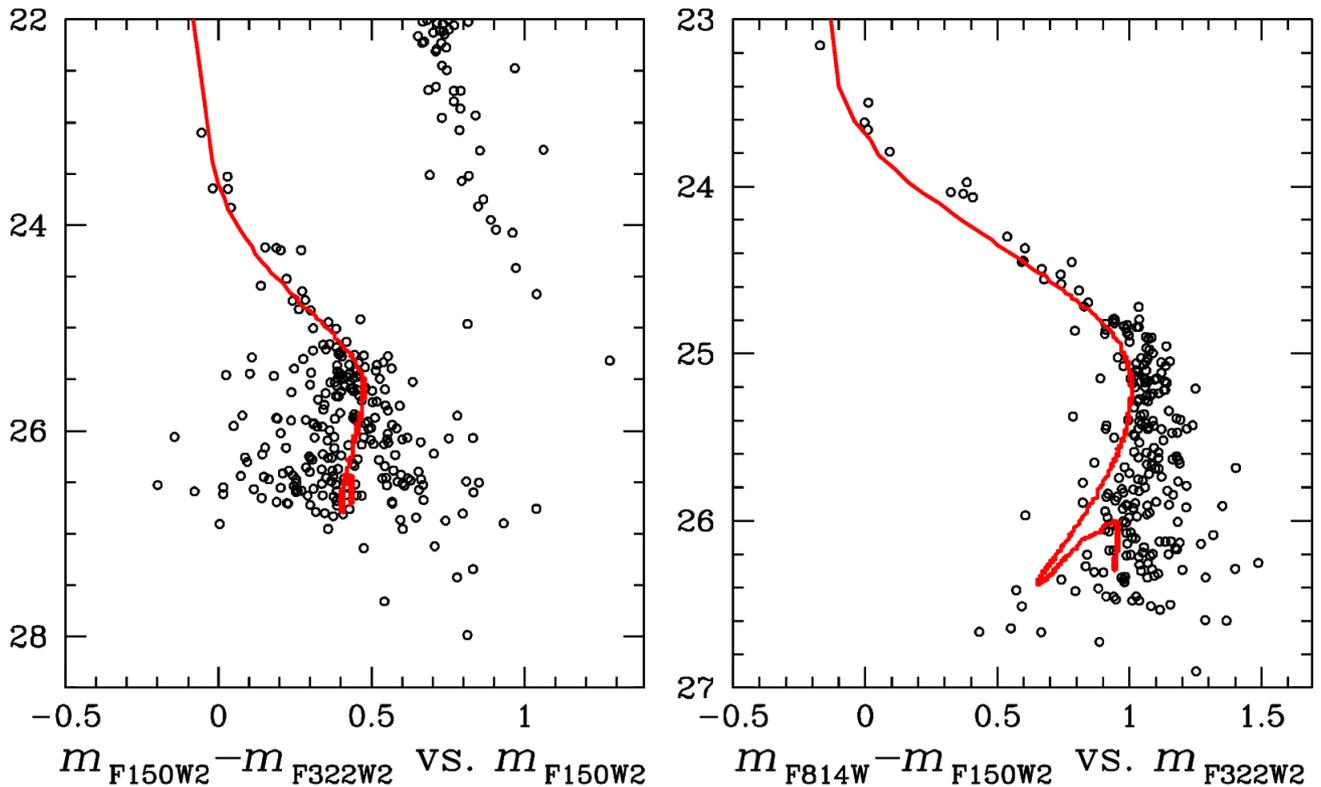

**FIGURE 5** The two most informative CMDs for the WD CS in NGC 6397, F150W2−F322W2 versus F150W2 (left), and F814W−F150W2 versus F322W2 (right). The tail of the MS can be seen in the upper right corner of the F150W2−F322W2 versus F150W2 CMD. The red lines trace a 13-Gyr DA-WD isochrone from Salaris et al. (2022).

where the completeness is considered reliable (Bedin et al., 2008).

The LF was then corrected for incompleteness using the completeness curve derived from the ASTs, and Poisson uncertainties from the observed counts were propagated using the completeness correction (thick line).

For a qualitative comparison with theory, we have calculated theoretical LFs using the 13 Gyr isochrone described above, calculating first a synthetic population in the appropriate filters that take into account the photometric errors. We proceeded as follows: Progenitor masses were first drawn from a power-law mass function (MF) $dN/dM \propto M^{\alpha}$, and the corresponding WD and its magnitude were determined by quadratic interpolation along the WD isochrone. We then added the distance modulus and extinction values to these magnitudes, and perturbed them randomly using Gaussian photometric errors as inferred from the ASTs.

To highlight the effect of the choice of the WD mass distribution along the cooling sequence (that in the simulation is determined by the convolution of the progenitor MF, the initial-final mass relation and the WD cooling times) we calculated LFs for $\alpha=-2.35$ (Salpeter, see Salpeter, 1955) and a top-heavy $\alpha=-1.6$.

The comparison with the observed LF shows that the chosen age is appropriate to match the faint end of the observed cooling sequence, that the exact shape of the theoretical LF (but not the magnitude of the faint cut-off) is obviously affected by the choice of the exponent of the progenitors' MF, and that even varying $\alpha$, the exact shape of the observed LF cannot be matched by theory.

The theoretical LFs display a pronounced dip in the range $25.7 \lesssim m_{F150W2} \lesssim 26.4$, irrespective of MF exponent, that is not present in the observational counterpart. This dip is caused by a flattening of the trend of increasing progenitor mass with increasing magnitude, and it is determined by the complex interplay between progenitor lifetimes, initial-final mass relation, and WD cooling times.

Clearly, the mass distribution along the observed cooling sequence can be further modified by dynamical evolutionary effects (e.g., selective mergers or ejection of low-mass progenitors or WDs) that are not accounted for in the simple modeling described above. As several *ad hoc* modifications could be considered to address the discrepancy between the detailed shape of the observed LF with the theoretical ones, we defer its analysis to a follow-up study.

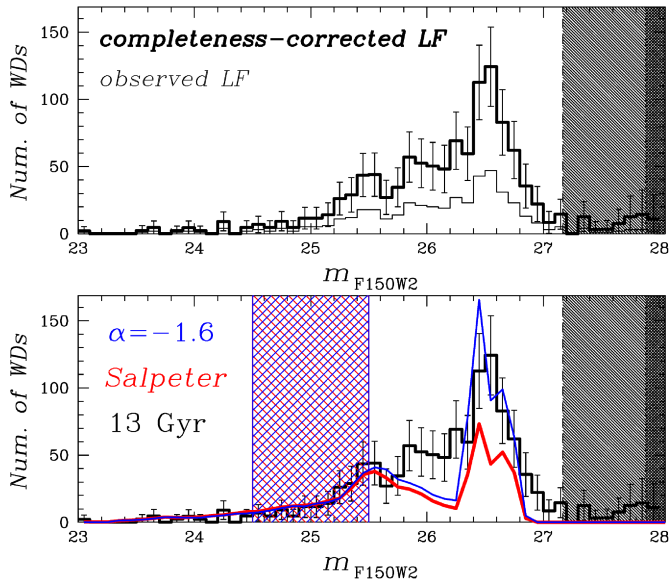

**FIGURE 6** *(Top:)* The LF derived for the NGC 6397 WD CS. The thin-line histogram shows the LF of the observed stars that passed all selection criteria, while the thick-line histogram shows the completeness-corrected LF with propagated Poisson error bars. Shaded areas indicate magnitudes where completeness drops below 50% (light grey) and 25% (dark grey). *(Bottom:)* Comparison of the observed WD CS LF to two theoretical LFs for an age of 13 Gyr, with different assumptions about the WD progenitor mass function (Salpeter in red, power-law exponent $\alpha = -1.6$ in blue). We neglect dynamical evolutionary effects that alter the resulting WD mass distribution. The shaded area on the left marks the magnitude range used for the normalization of the theoretical LFs.

### 4.3 | CIA vs. high masses WDs

In the "classic" *HST*/ACS/WFC CMD F606W−F814W versus F814W, which offer the most effective filter combination for *HST* surveys (as it optimizes both depth and field coverage; e.g., Bedin et al., 2008, 2004, 2015, 2023, 2010; Kalirai et al., 2012; Richer et al., 2006), there is a blueward turn at the bottom of the WD isochrone that corresponds to increasingly more massive, and therefore smaller radii, WDs that originate from shorter-lived progenitors. This population also comprises the LF peak observed in our sample. However, this effect in the optical CMD is *degenerate* with another effect that *also* move the WD CS blue ward. Indeed, depending on the cluster age, the blueward turn is further enhanced by the onset of collision-induced absorption (CIA) from molecular hydrogen in cool WD atmospheres (Hansen, 1998). The strong CIA bands reduce the infrared flux and redistribute it to shorter wavelengths, producing an enhancement of the blueward turn of the CMD sequence with decreasing $T_{\rm eff}$ and luminosity.

In the NIRCam filter F150W2, the CIA effect begins at a higher $T_{\rm eff}$, when the isochrone coincides with the cooling track of lower-mass WDs. The effect of CIA in this filter is therefore clear and unambiguous, without overlap with the blueward turn produced by higher-mass, hotter WDs, which become significant only at fainter magnitudes. In the *JWST* infrared bands, the WD isochrone displays a blueward turn due to CIA $H_2$ at $m_{\rm F322W2} \sim 25$ corresponding to $T_{\rm eff} \sim 5000$ K, down to $\sim 26$. Indeed, between $24.5 \lesssim m_{\rm F322W2} \lesssim 25.5$ the WD mass along the isochrone changes by only $\Delta \mathcal{M} = 0.03\,M_\odot$ at an average mass of $\mathcal{M} = 0.56\,M_\odot$; i.e., across this magnitude range there is less than a 5% difference in mass. The observed cooling sequence displays the same turn to the blue in the same magnitude range predicted by theory, showing that the current modeling of CIA $H_2$ adequately reproduces the trends in the observations, if not the exact colors.

### 4.4 | A possible Red excess in WDs

The CMD F150W2−F322W2 versus F322W2 in Fig. 5 shows hints of an unexpected feature. The lowest part of the WD CS is considerably broader than expected given the photometric errors. Keeping in mind that our errors could be underestimated, and that models are far from representing perfectly the location of the bottom of the WD CS in these CMDs, in this section we speculate on the possibility that the observed broadening is intrinsic and not entirely due to photometric errors. Therefore, the distribution suggests two distinct groups of objects between $m_{\rm F150W2} \sim 25.5$ and 27 separated at $m_{\rm F150W2} - m_{\rm F322W2} \approx 0.5$. Even more interestingly, while the blue component seems to approximately follow the shape of the model isochrone, the red component diverges to the red by up to 0.5 mag in color at the faintest magnitudes.

Given that this behavior is unexpected, we carried out additional tests to verify this result. As a first step, we checked whether the WDs at the bottom of the CS are also separated in color in other CMDs. Figure 7 displays nine CMD combinations using the four *HST* and *JWST* filters, with the blue and red WDs in the *JWST* explicitly red/blue color-coded. Generally, these two groups are *not* separated in the CMDs constructed from combinations of the *HST* filters and the *JWST* F150W2 filter. However, all CMDs that include the *JWST* F322W2 filter *do* show a clear separation between the two WD groups in color and/or magnitude. Of particular interest are those CMDs with the F322W2 magnitude on the vertical axis, where the red WD group is on average the magnitude and color of the stars within each bin. We then smoothed each fiducial line with a cubic spline, and "rectified" the CMD, following the methodology described in Libralato et al. (2019). Briefly, we defined the quantity

$$\Delta(m_{\text{F150W2}} - m_{\text{F322W2}}) = ((m_{\text{F150W2}} - m_{\text{F322W2}}) - \text{fiducial}_{\text{blue}})/$$
$$(\text{fiducial}_{\text{red}} - \text{fiducial}_{\text{blue}})$$
(1)

where fiducial$_{\text{red/blue}}$ correspond to the color of a star at a given $m_{\text{F150W2}}$ if placed along the red/blue fiducial lines. Finally, we performed a double-Gaussian fit to the distribution of sources across $\Delta(m_{\text{F150W2}} - m_{\text{F322W2}})$ (i.e., marginalizing over the $m_{\text{F150W2}}$ magnitude) to obtain a tentative estimate of the fraction of WDs in the two groups. We again followed the methodology described in Libralato et al. (2019) to remove the dependency on the bin width and the starting point of the color histogram. We computed this histogram 1 000 times, each time adding a random Gaussian offset corresponding to the photometric error of the star, and averaged these histograms to create our final color distribution. From this fit, we estimate that the *blue* group accounts for 74 ± 4% and the *red* group accounts for 26 ± 4% of the WDs at the bottom of the CS.

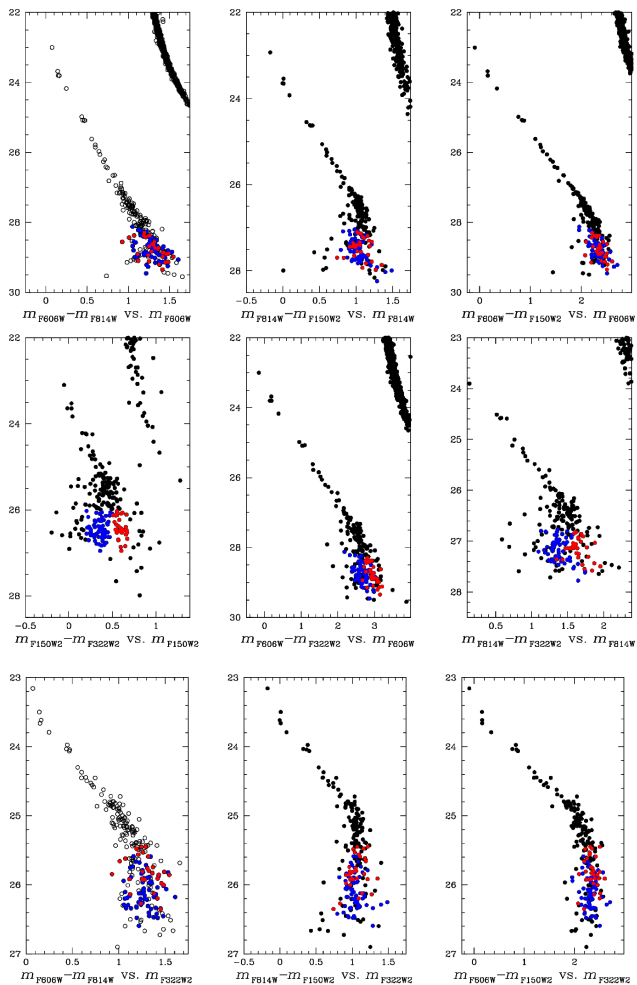

**FIGURE 7** Nine CMD combinations from the four *HST* and *JWST* filters employed in this study. These figures highlight the red and blue WD populations originally distinguished in the F150W2−F322W2 versus F322W2 CMD (leftmost panel in the middle row).

As an attempt to explain the observed excess of flux in the F322W2 filter for a subsample of faint WDs, we explored a simple model in which we assumed that the red WDs are composed of WD+BD binaries. We constructed two synthetic populations of binaries for this model: WD+WD pairs assuming random-mass components, and WD+BD pairs assuming a range of BD masses. We first mapped a continuous distribution of single WDs with varying magnitudes running parallel along the expected WD CS based on the Salaris et al. (2022) isochrone. We constructed WD+WD pairs from the isochrones assuming random WD-mass for the two components. For the BD companions, we extended the isochrones of Gerasimov et al. (2024) computed for the case of NGC 104, into the BD regime, and used these to calculate the CMD locations of hypothetical WD+BD binaries over a BD mass range of 0.046 $\mathcal{M}_\odot$ and 0.080 $\mathcal{M}_\odot$. The middle panel of Figure 9 shows how these WD+BD combinations produce a continuum of reddened sources reaching the MS track for the most massive BD companions, which would dominate the combined light of the system. Notably, at $m_{\text{F150W2}} \approx 25$ a split emerges between the single WD track and the WD+BD track for the lowest BD companion masses that roughly aligns with the split between the blue and red WD populations. Indeed, an ad-hoc model that shifts all of our observed WDs to the WD CS isochrone, assumes a 5% WD binary fraction and a 25% WD+BD binary fraction, with BD companion masses drawn from a uniform distribution between 0.04 $\mathcal{M}_\odot$ and

$\Delta m_{\text{F322W2}} = 0.5$ mag more luminous than the blue WD group.

However, if the photometric errors in F322W2 are underestimated the overall scatter might be explained just by random colour fluctuations (more below).

As a next verification step, we quantified the fraction of WDs in each of these two groups, as illustrated by Fig. 8 . We first fit fiducial lines to the blue and red WDs, defined as sources with $m_{\text{F150W2}} > 25$ and with ($m_{\text{F150W2}} - m_{\text{F322W2}}$) smaller or larger than 0.5, respectively. Each line was obtained by binning the data in 1 mag bins in steps of $\Delta m_{\text{F150W2}} \sim 0.2$ mag each, and computing the 3$\sigma$-clipped median value of $0.06\,\mathcal{M}_\odot$, closely resembles the observed dual tracks of blue (single) and red (binary) WDs. Note that the simulated single WD sequence (in magenta) in the right-hand panel of Fig. 9 , is not able to reproduce the breadth of the observed faint end of the WD CS, regardless of the exact location on the simulated WD CS in the CMDs.

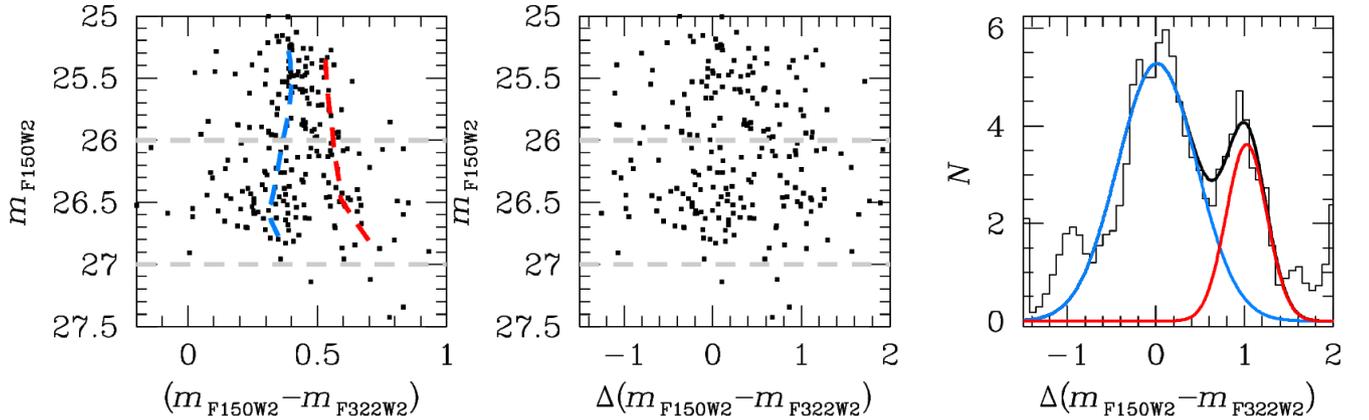

**FIGURE 8** Illustration of our procedure for estimating the fraction of sources that correspond to the two putative branches of the lower part of the WD CS. The left panel shows the F150W2−F322W2 versus F322W2 CMD with blue and red lines tracing the fiducial trends for objects with ($m_{F150W2} - m_{F322W2}$) less than or greater than 0.5, respectively. The two horizontal dashed lines delimit the region where the two populations are best-separated. The center panel shows the "rectified" CMD based on the quantity defined in Eqn 1. The right panel shows a double-Gaussian fit to the rectified CMD following Libralato et al. (2019), with the blue and red histograms corresponding to the blue and red branches of the WD CS.

While this simple model reproduces the observed WD red excesses, the high WD+BD binary fraction required greatly exceeds the rate of BD companions to WDs detected in the local Galactic disk, which is reported to be ≲0.5% (Farihi, Becklin, & Zuckerman, 2005; Rebassa-Mansergas et al., 2019; Steele et al., 2011). The multiplicity demographics of WDs in NGC 6397 would have to be substantially different from the local disk for our toy model to explain the observed excess. An alternative hypothesis would be the presence of dusty material around these WDs, in the form of tidally disrupted debris disks (Chary, Zuckerman, & Becklin, 1999; Reach et al., 2005). However, the cooler temperatures of the evolved WDs at the end of the CS (T < 5,000 K) would likely result in a disk too cold to be detectable. Reddening could also arise from residual dust from the post-asymptotic giant branch (post-AGB) phase, although these would have likely since long dissipated for the massive WDs at the end of the CMD, and moreover only about 3% of local WDs show evidence of IR excess attributable to dust (Barber, Kilic, Brown, & Gianninas, 2014).

Interestingly, the high-precision multi-band *HST* photometry by Milone et al. (2012a) has revealed that the MS of NGC 6397 splits into two components, a "primordial" population with chemical abundance ratios similar to the Galactic disk comprising ~30% of stars, and a "second generation" of stars with enhanced He, Na and N and depleted C and O comprising ~70% of stars. It is tempting to associate the two putative WD populations identified in this study with two MS populations previously detected in NGC 6397, with the red WD group associated with the primordial population and the blue WD group associated with the second generation population. However, this is a speculative claim as there is insufficient data and modelling to explore the correlation between these two populations.

Other scenarios can also be explored. Indeed, in our isochrone analysis we did not account for populations of He-dominated atmosphere WDs, or low-mass He-core objects. However, both possibilities do not appear viable, as discussed below.

Figure 10 compares in two different CMDs the cluster cooling sequence and representative $0.5\,\mathrm{M}_\odot$ WD tracks with atmospheres of pure He, pure H and mixed H/He (with log H/He = −2, where H and He denote number fractions) respectively, from Caron, Bergeron, Blouin, and Leggett (2023) calculations.

In the F150W2−F322W2 versus F322W2 CMD, which shows the population of WDs with red excess (see also Fig. 7 ), the tracks with helium-dominated atmospheres are bluer than hydrogen-atmosphere ones, not redder. We can hypothesize that due to uncertainties in the Lyα wing opacity, the H⁻ opacity, or the CIA opacity, or a mix of those (see

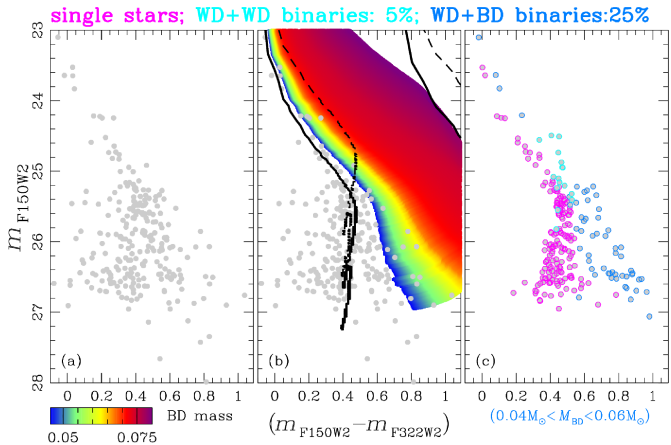
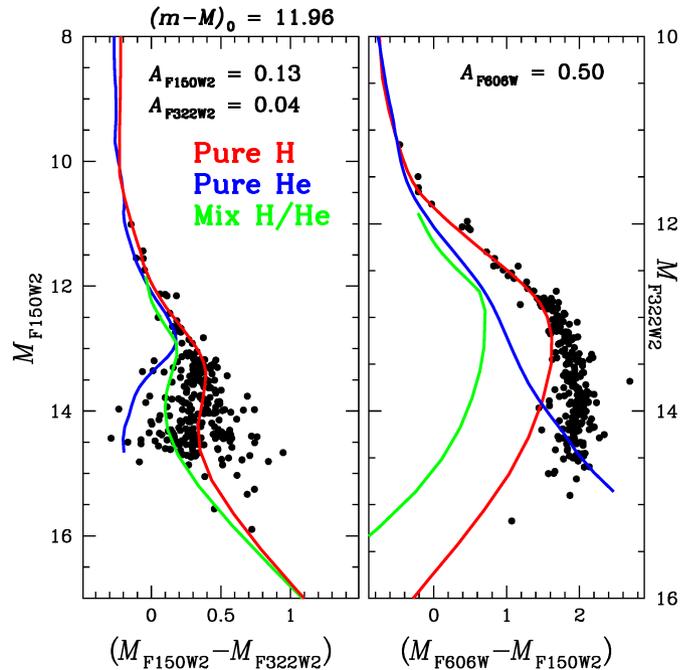

**FIGURE 9** A toy model of the WD CS of NGC 6397 assuming a 25% fraction of WD+BDs binaries and 5% fraction of WD+WD binaries. The left panel shows is the observed CMD in the WD regime (cf. Figure 4 ). The middle panel overplots isochrones for WDs and MSs (the latter based on Gerasimov et al. 2024), and the equal-mass loci of WD+WD and MS+MS binaries (black dashed lines). The shaded regions show the binary trend lines for WD+BD binaries as a function of BD mass, ranging from 0.046 $\mathcal{M}_\odot$ to 0.080 $\mathcal{M}_\odot$ (color bar). The right panel shows a simulated population of single (magenta circles) and binary WDs assuming a 5% fraction of WD+WD binaries (light blue circles) and a 25% fraction of WD+BD binaries with BD masses drawn between 0.04 $\mathcal{M}_\odot$ to 0.06 $\mathcal{M}_\odot$ (dark blue circles).

Caron et al., 2023) the H-atmosphere models and isochrones are actually redder than our adopted calculations, and can match the redder branch of the cooling sequence in this CMD, whilst models with He-dominated atmospheres match the bluer bulk of the cluster WDs.

However, this explanation appears in disagreement with the second CMD in Fig. 10 , where no red excess objects appear (see also Fig. 7 ). In this CMD models with He-dominated atmospheres are again well separated in colour from the H atmosphere counterpart, at odds with the data.

In the case of He-core WDs, due to their lower mass (hence larger radii) compared to the CO-core counterparts, they would appear redder than the main cooling sequence, potentially explaining the red excess objects in the F150W2−F322W2 versus F322W2 CMD. However, larger radius He-core WDs are redder than CO-core WDs also in the other CMDs shown in Fig. 7 , again in disagreement with observations.

To find a definitive explanation for the red excesses observed in the WD CS of NGC 6397 will likely require both further observational follow-up, particularly with longer wavelength filters, to assess the spectral characteristics of the observed excess, and further theoretical explorations to identify testable observational predictions.

**FIGURE 10** Two different CMDs of the cluster cooling sequence, compared to representative 0.5 $M_\odot$ WD tracks with pure He, pure H and mixed H/He compositions (see text for details). The observed cooling sequence has been displayed as absolute magnitudes versus dereddened colours using the labelled distance modulus and extinctions.

## 5 | CONCLUSIONS

We have presented the first high-precision, multi-filter photometry and astrometry of the globular cluster NGC 6397 extending to the near-infrared bands through deep *JWST*/NIRCAM imaging. Our dataset provides a comprehensive view of the entire WD population of the cluster. Our main findings are as follows:

- Our *JWST*/NIRCAM F150W2 and F322W2 photometry encompasses a roughly 10 arcmin$^2$ field of view about 6 arcmin from the NGC 6397 cluster core, and reaches magnitudes of $m_{F150W2} = 27.17$ and $m_{F322W2} = 26.88$ on the Vega magnitude scale, well below the apparent termination of the WD CS.

- Taking advantage of the ultradeep *HST*/ACS/WFC optical imaging obtained ∼18 yr prior to the *JWST* observations, we have been able to make an accurate proper motion selection of NGC 6397 cluster members to a relative precision of 0.3 mas yr$^{-1}$, excluding with high confidence background sources from the extended Bulge and outside the Milky Way.

- Our data reveal the entirety of the WD CS, which aligns well with DA WD cooling models. In particular, the infrared data allow us to separate two color trends at the faint end of the CS that are intermingled in optical CMD: the onset of CIA $H_2$ absorption in low-mass, low-temperature WDs, which shifts the isochrone to blue in $m_{F814W} - m_{F150W2}$ colors; the contribution of massive WDs at the lowest luminosities along the CS.

- We have determined the WD infrared LF corrected for the sample completeness. The magnitude of the observed faint cut-off is matched by theoretical LFs for an age of 13 Gyr. The exact shape of the observed LF is not reproduced by theoretical LFs calculated with power-law progenitor MFs. An accurate match of the whole observed LF very likely requires to take into account dynamical evolutionary effects that affect the WD mass distribution in the observed field.

- We find hints of a distinct population of WDs showing excess flux in F322W, comprising roughly 25% of sources $m_{F150W2} > 25.5$. While this putative feature deserves further confirmation, this excess is limited to F322W2, and broadens toward fainter magnitudes. We attempted several explanations and scenarios for this excess, including a large population of WD+BD binaries, reddening from circumstellar debris disks or post-AGB material, contributions from helium-dominated atmosphere and helium-core WDs, and a potential tie to the two populations present in the stellar Main Sequence of NGC 6397. We have not found any clear explanation for this excess.

This study provides our first deep view of the NGC 6397 system at infrared wavelengths, and subsequent studies will take advantage of this data to probe the MS population into the substellar regime, field populations around the cluster, and cluster kinematics. Future *JWST* epochs will allow us to extend proper-motion membership to fainter objects and well into the BD sequence. Extending these observations to other long-wavelength NIRCam filters (F277W, F356W, F444W) and possibly MIRI at 10 $\mu$m (F1000W) will allow us to confirm and –in this case– spectroscopically characterize the infrared excess detected around a significant fraction of NGC 6397 WDs, disentangling for instance sub-stellar companions from dust emission (Reach et al., 2005; Swan et al., 2023).

In addition, further theoretical investigation would be needed to better explore the nature of the apparent observed red excess as well as the evolution of the cluster MS. To this aim, we provide our reduced data, high-accuracy photometric catalogue for members, and atlases in multiple filters as part of this article's supplementary on-line material.

## 6 | ACKNOWLEDGMENTS

We thank an anonymous Referee for the prompt review of our manuscript, and for the useful suggestions. We warmly thank STScI, our Program Coordinator and Instruments Reviewers –Shelly Meyett, Mario Gennaro, Paul Goudfrooij and David Golimowski– for their great support during the review of our problematic observations. LRB, DN, MG and MSc acknowledge support by INAF under the WFAP project, f.o.:1.05.23.05.05. MS acknowledges support from The Science and Technology Facilities Council Consolidated Grant ST/V00087X/1. ABu, DA, JA, RG, and ABe, acknowledge support from STScI funding associated with GO-1979.

## APPENDIX A: SELECTION PROCEDURES AND COMPLETENESS

Figure A1 illustrates the selection procedures on the initial local maxima detected by `KS2` in both F150W2 and F322W2 *JWST*/NIRCAM images, a total of 43,040 sources. To incorporate the widest temporal range, we only selected sources that fell in the common area between the *JWST* and *HST* data sets, a total of 23737 sources. We further narrowed our sample down to those sources for which the S/N would allow detection above the local sky background, the latter determined from the `logRMSsky` quantity that measures the local sky noise at a given location (see Bedin et al., 2008, 2009). This criterion reduced the number of bona-fide sources to 6171; it also removed the vast majority of the extended sources. We then conservatively defined by eye a region on the CMD that is expected to contain cluster WDs, indicated by the green lines in Fig. A1. Then, instead of using `RADXS` to select the best measured stars, we used proper motion membership as defined in Sect. 3.1 to isolate cluster members, leaving 459 sources. Finally a generous selection on the PSF quality-of-fit parameter q (see Anderson, King, et al., 2008) removed 2 sources, resulting in 457 robust WD cluster members. This is the sample used in our analysis of the WD Luminosity Function.

When comparing the theoretical isochrones to the observed CMDs, we performed a second, tighter selection, retaining only the best measured stars. In this case, we imposed a selection on the stellarity parameter `RADXS` (Bedin et al., 2009) with the result shown in Fig. A2.

Since this parameter cannot be computed for about 20% of the faintest sources detected by `KS2`, it significantly reduces completeness and thus compromises the statistical robustness of the WD CS LF computation. However, for our CMD analysis a clean CMD was considered more important than completeness at the faintest magnitudes. The completeness for the two cases are illustrated in Fig. A3. Finally, we release both sample of WDs stars, with (457 objects) and without (245 objects) the `RADXS` selections.

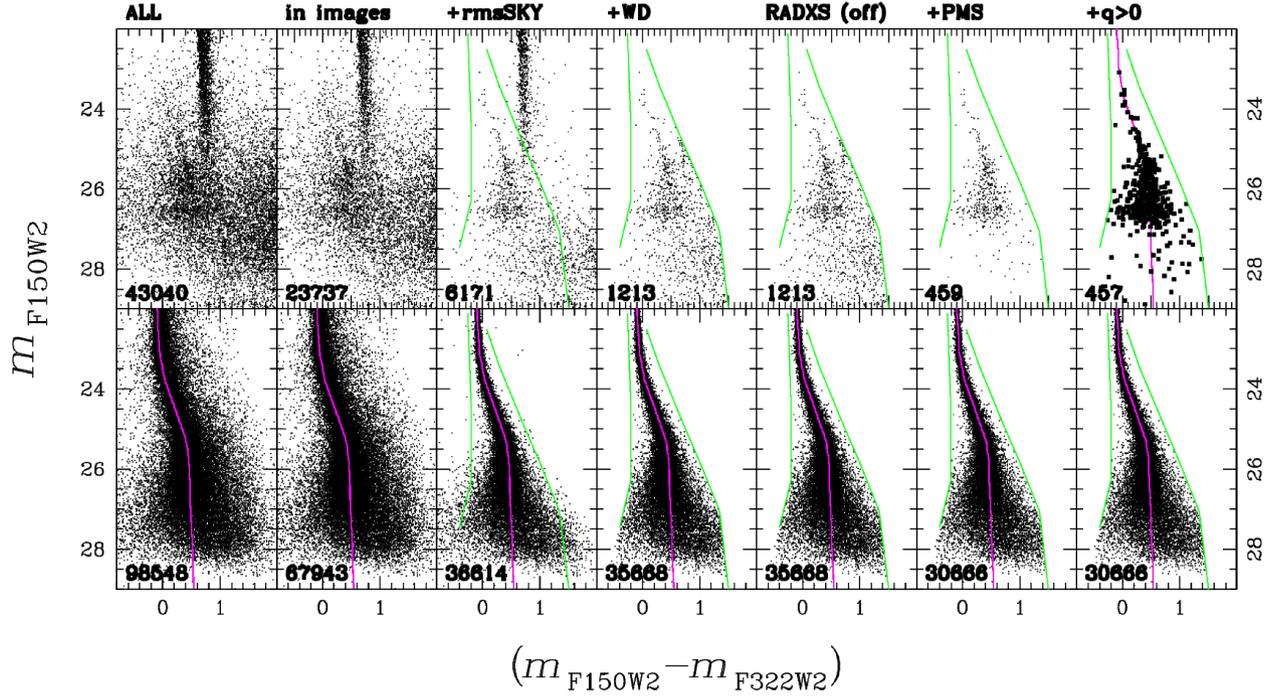

**FIGURE A1** A summary of the selection process that goes from reducing all local maxima, to a sample of WD with which the WD LF and its completeness correction are computed. Panels in the top row, show from left to right the selection progression for *real* stars, while the row of panels below the one for *artificial* stars. In this figure no selection was made on RADXS. Green lines delimit the CMD regions within which member sources are assumed to be WDs. In magenta the input magnitudes for ASTs (see text).

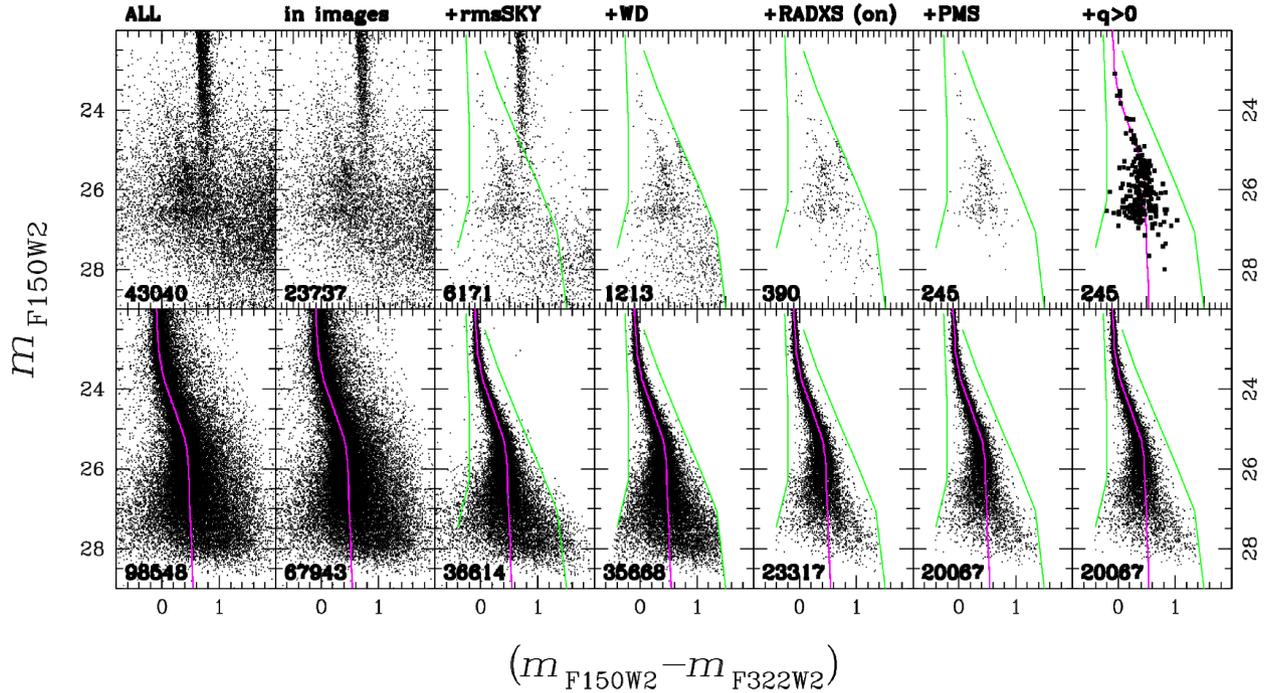

**FIGURE A2** Same as for Fig. A1 , but with at selection on RADXS.

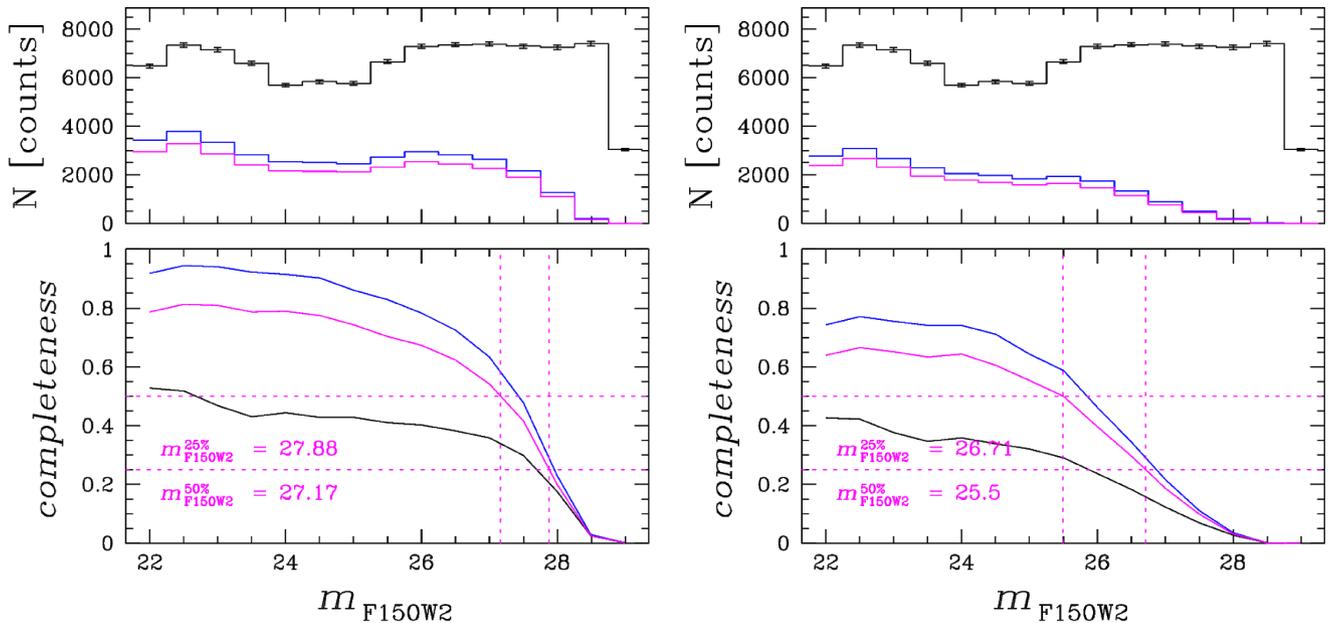

**FIGURE A3** *(top panels):* the histogram of the counts for: the added stars (black), for the recovered stars on suitable "good" region (blue) and for those among these that also passed the PMs membership (magenta). *(bottom panels):* the completeness curves: the overall completeness $c$ (black), the completeness on the "good" region (blue), and the one for sources that also passed the PMs membership criterion $c_g$ (magenta). Dotted horizontal lines mark the 25% and 50%-level of $c_g$, while the vertical lines and the labels give the values. *(Left panels):* without selections on `RADXS`. *(Right panels):* with selections on `RADXS`.